# Bright continuously-tunable VUV source for ultrafast spectroscopy

Lucie Jurkovičová[1,2], Ltaief Ben Ltaief[3], Andreas Hult Roos[1], Ondřej Hort[1], Ondřej Finke[1,2], Martin Albrecht[1,2], Ziaul Hoque[1], Eva Klimešová[1], Akgash Sundaralingam[3], Roman Antipenkov[1], Annika Grenfell[1], Alexandr Špaček[1,2], Wojciech Szuba[1], Maria Krikunova[1,4], Marcel Mudrich[3,+], Jaroslav Nejdl[1,2,*] & Jakob Andreasson[1]

Ultrafast electron dynamics drive phenomena such as photochemical reactions, catalysis, and light harvesting. To capture such dynamics in real-time, femtosecond to attosecond light sources are extensively used. However, an exact match between the excitation photon energy and a characteristic resonance is crucial. High-harmonic generation sources are exceptional in terms of pulse duration but limited in spectral tunability in the VUV range. Here, we present a monochromatic femtosecond source continuously tunable around 21 eV photon energy utilizing the second harmonic of an OPCPA laser system to drive high-harmonic generation. The unique tunability of the source is verified in an experiment probing the interatomic Coulombic decay in doped He nanodroplets across the He absorption bands. Moreover, we achieved intensities sufficient for driving non-linear processes using a tight focusing of the VUV beam. We demonstrated it on the observation of collective autoionization of multiply excited pure He nanodroplets.

[1]ELI Beamlines Facility, The Extreme Light Infrastructure ERIC, Za Radnicí 835, 25241 Dolní Břežany, Czech Republic. [2]Czech Technical University in Prague, FNSPE, Břehová 7, 115 19 Prague 1, Czech Republic. [3]Institute of Physics and Astronomy, Aarhus University, Denmark. [4]Technical University of Applied Sciences, Hochschulring 1, 15745 Wildau, Germany.
*email: Jaroslav.Nejdl@eli-beams.eu, +email: mudrich@phys.au.dk

Observation and control of ultrafast electron and nuclear dynamics in atoms, molecules, and extended systems using ultrashort light pulses are highly important for advancing atomic and molecular physics, physical chemistry, and material sciences. Recent developments of VUV ultrafast pump-probe schemes have brought fascinating discoveries, like identifying pathways of photo-triggered chemical reactions [1, 2, 3], tracking electron correlations in real-time [4, 5, 6], or controlling the optical properties of semiconductors [7]. At the same time, through their ability to deliver exceptional intensity on target, laser sources drive the research into non-linear phenomena based on high-intensity light-matter interactions. High-order harmonic generation (HHG) in gases driven by femtosecond laser pulses is a well-established technique enabling tabletop sources of femto- and attosecond VUV to soft X-ray radiation [8, 9, 10, 11, 12]. These sources traditionally exhibit outstanding properties in terms of ultra-short pulse duration, coherence, and polarization control. However, as the frequencies of the comb of generated harmonics are directly related to the wavelength of the driving laser, continuous spectral tunability is not straightforward to achieve.

As many applications require VUV energies in resonance with specific excitations in the sample [4, 5, 13, 14, 15, 16, 17] to initiate a distinct reaction pathway or decay process, several approaches to realize spectral selectivity or tunability have been developed. For specific applications, grating monochromators [18, 19] that allow individual harmonics to be selected over a wide range of energies are used. A grating monochromator has limited throughput and is often used in a forward-focusing geometry that yields a large focal spot resulting in limited intensity on target. For applications where continuous tunability is necessary, other ways to realize spectral tuning of HHG sources have been investigated, e.g., using optical parametric amplifiers [20], controlling the laser pulse chirp [21], mixing two different driving wavelengths [22], controlling the nonlinear propagation of the driving laser pulse [23, 24, 25] or using double pulses [26]. Recently, a broadband optical parametric chirped-pulse amplifier (OPCPA) laser system has been used to generate tunable pulses between 25 to 50 eV [27]. However, this approach has so far been limited to a high repetition rate and low-intensity applications. To our knowledge, no continuously tunable HHG source with a focused intensity sufficient for non-linear VUV light-matter interactions has been reported.

Achieving high flux and high intensity on target with an HHG source is challenging. Even under ideal conditions, the conversion efficiency of the HHG process is typical of the order of $10^{-5}$. To our knowledge, the highest reported conversion efficiency to the VUV region is of the order of $10^{-4}$ reaching a photon flux of $10^9$ photons per pulse (at the source) at ≈1 MHz repetition rate [28, 29]. Indeed, several concepts for generating high-flux single harmonic beams in the VUV region (up to $10^8$ photons/pulse) have been reported [30, 31], however, without the ability of spectral tuning. At lower repetition rates, VUV intensities from HHG of the order of $10^{14}$ W/cm$^2$ have been reported [32, 33, 34, 35] showing that an HHG beamline, when optimized for this, can reach intensities sufficient for highly non-linear VUV science. For these high-intensity applications, back-focusing multilayer (ML) optics with energy-selective reflectivity [36, 37] are often used to achieve tight focusing of the HHG beam. This approach minimizes the number of optical elements in the VUV beam propagation and maintains the ultrashort pulse duration from the source. However, it has been reported that difficulties in focusing all the harmonic orders in one focal spot may significantly decrease XUV intensity on the target [38, 39].

In this work, we present an approach to HHG and VUV beam focusing that simultaneously meets the requirements for continuous spectral tunability and monochromaticity of the high-harmonic source while preserving high VUV flux out of the source and reaching an intensity on target above the threshold for non-linear many-body processes. This combination of parameters has until now only been reported for seeded free-electron lasers (FELs) [40, 41, 42]. We demonstrate the spectral tunability and intensity of our source through measurements of ion time-of-flight (TOF) mass spectra and electron velocity map imaging (VMI) spectra of resonantly excited pure and doped helium nanodroplets.

This is the first time, to our knowledge, that the unique features of the OPCPA laser technique have been used to generate a continuously tunable VUV beam in a high-intensity application.

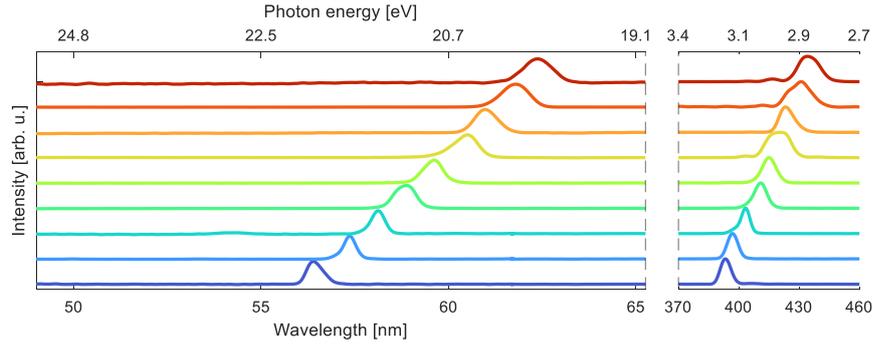

**Fig.1 Continuously tunable experimental HHG spectra within >2 eV range**. The spectrum of the 7[th] harmonic (left panel) tunable in the range from 20 eV to 22 eV generated by the second harmonic (right panel) of the IR laser in a Kr gas target. The spectra are vertically offset for clarity.

**Results**
**Generation of spectrally tunable monochromatic high-flux VUV pulses: Experimental implementation.**
We present a method for generating high-order harmonics in the VUV region with photon energy that is continuously tunable over a 2 eV range (Fig. 1). To achieve this, we drive HHG in krypton gas with the second harmonic (SH) of the high-intensity IR beam from the in-house developed *Allegra* broadband OPCPA laser system [43]. The experimental setup is shown in Fig. 2. For the experiment, the *Allegra* laser was set to deliver pulses at a 1 kHz repetition rate with a pulse energy of 20 mJ compressible down to 13 fs and a broadband spectrum spanning between 750 nm and 930 nm (shaded area of Fig. 3). The central wavelength of the OPCPA can be tuned by a slight adjustment of the pump delay relative to the amplified chirped pulse. In addition, an adjustment of the group delay dispersion (GDD) in the presence of a small amount of third-order dispersion (TOD) of the IR pulse allows tuning of the central wavelength of the second harmonic generation (SHG) output. For the SHG, we used a BBO crystal optimized for high conversion efficiency while preserving the short pulse duration (see Methods). Wide-range spectral tuning of the SHG is then achieved by changing the phase-matching in the conversion crystal combined with adjusting the amplification of the OPCPA laser system (see Methods). The spectra of the SH beam used to drive the HHG for various angles of the frequency-doubling crystal are shown in Fig. 1 (right panel) together with the corresponding spectra of the generated 7[th] harmonic (left panel).

We note that the bandwidth of the SH spectrum increases with the central wavelength from 140 meV on the blue side to 280 meV on the red side (Fig. 1 right panel). The expected pulse duration of the SH driving pulses should always be shorter than 30 fs. The bandwidths of the SH pulses measured experimentally are in full agreement with our numerical modeling of SHG in the BBO crystal. In Fig. 3, the calculated phase-matching curves are shown for two extreme cases featuring similar bandwidths to the ones observed experimentally in Fig. 1. The phase-matching bandwidth of the BBO crystal was significantly narrower than the spectral width of the primary laser pulses (see Fig. 3), which provided the opportunity to change the central wavelength of the SHG by tilting the crystal (θ). Note that SHG of broadband pulses can be considered both as a combination of sum-frequency generation of spectral sidebands and pure SHG. This results in a high conversion efficiency even if the phase-matching bandwidth is narrower than the laser spectrum, allowing for efficient continuous spectral tuning in a wide range. The generated UV radiation was spectrally cleaned from the residual IR component by a thin dichroic mirror and focused on the Kr gas target for HHG. The generated VUV beam is then filtered from residual driving laser radiation by a set of aluminum metallic filters and either characterized by a wavelength-calibrated flat-field spectrometer or sent to the user end-station [44] (see Fig. 2).

In the application chamber (the Eli Beamlines MAC station [44]) the VUV beam is back-focused by a broadband multilayer spherical mirror with a focal length of 75 mm at normal incidence. Using the SH to drive the HHG results in a 6 eV energy separation between adjacent harmonic orders. This can allow us to efficiently monochromatize the VUV beam employing a multilayered focusing mirror with a coating chosen so that the 9[th] harmonic order is outside of the reflectivity band and the 5[th] harmonic order is blocked by a 200 nm thin aluminum filter in the VUV beam path (see Methods). As a result, we can obtain a monochromatized VUV beam without a grating monochromator.

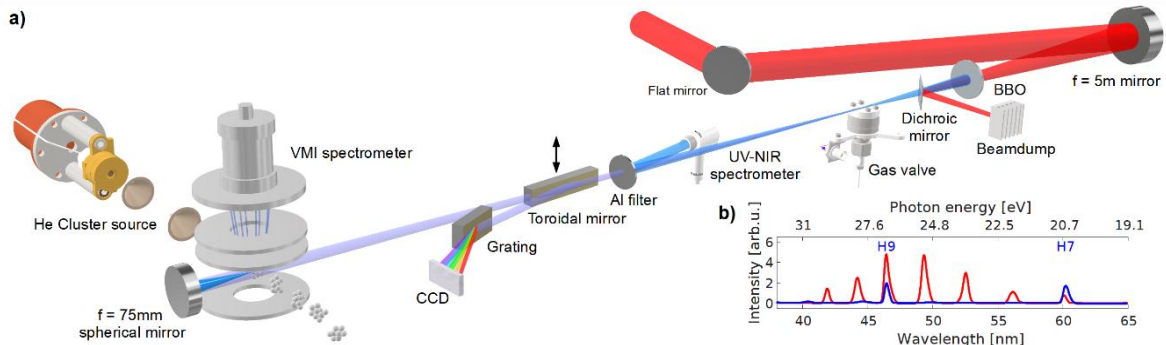

**Fig. 2 Experimental setup. a.** The experimental setup for tunable HHG implemented into the HHG beamline at ELI Beamlines [8]. The IR laser beam is apertured by a motorized iris and focused by a spherical mirror at almost normal incidence on a Kr gas target. A BBO crystal in the beam path is used for the second harmonic generation, and a thin dichroic mirror is used to separate SH from IR. Thin metallic filters separate the generated VUV from residual pump light. The partially reflected pump beam from the metallic filters is characterized by a UV-NIR spectrometer. A wavelength-calibrated flat-field CCD spectrometer can be used to characterize the VUV beam, or the beam is sent to the user end-station MAC. The MAC station [44] is set up to investigate He nanodroplets, including a cryo-cooled pulsed He source, heated cells for doping the droplets with evaporated substances, and a velocity map imaging detector for either electrons or ions detection. **b**. Typical spectra obtained when VUV pulses are generated by HHG from both the fundamental and SH as a drive, i.e., without the dichroic mirror (red line), and when only the SH pulses are used to drive HHG, i.e., with the dichroic mirror (blue line).

The overall conversion efficiency from IR laser pulses via SHG to the single 7$^{th}$ harmonic pulses was measured to be approximately $10^{-5}$ in the source region with a VUV beam divergence of 0.6 mrad. The single-harmonic photon flux was measured to be $2\times10^7$ photons/pulse on target (corresponding to a VUV pulse energy of about 80 pJ). An upper limit of the intensity achieved on the target is obtained from simulations of the VUV beam generation and propagation through the beamline optics and is in the order of $7\times10^{11}$ W/cm$^2$ (see Methods).

The target consists of a pulsed beam of pure or Li-doped He nanodroplets interacting with the HHG beam at the center of a VMI detector (see Methods), which was used to acquire electron spectra or ion TOF traces from the VUV-irradiated nanodroplets.

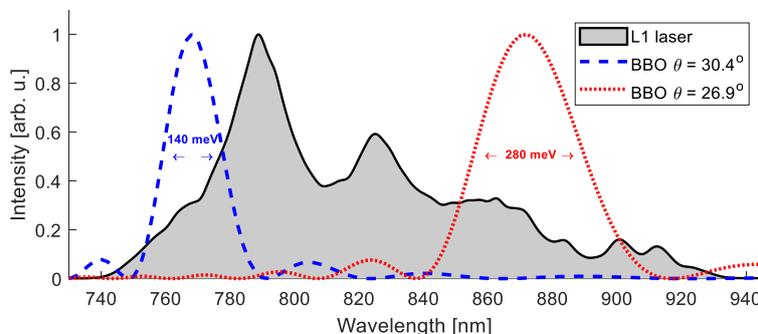

**Fig. 3 Spectrum of *Allegra* laser & phase-matching of BBO crystal.** Black line: A typical IR spectrum produced by the OPCPA laser system. Blue and red lines: Two phase-matching curves of the BBO crystal calculated numerically for two extreme values of the phase-matching angle θ used during tuning. Second harmonic generation can be achieved in the whole range between these two curves by continuously tuning the phase-matching angle θ.

**VUV spectroscopy of He nanodroplets by tuning a single harmonic.** The scientific relevance of our continuously tunable high-intensity VUV source is demonstrated in a proof-of-principle experiment on He nanodroplets doped with alkali metal atoms (Li). By resonant excitation of He nanodroplets below the ionization potential of He ($E_i \approx 24.6$ eV), we have studied their autoionization by interatomic Coulombic decay (ICD), which is traditionally termed Penning ionization [45]. In this process, the photon-energy-dependent absorption of one VUV photon by a He atom creates a He* excited state within the He nanodroplet. The absorbed energy is then transferred from the photo-excited He* to a Li atom attached to the surface of the same He nanodroplet, thereby producing a Li$^+$ ion and an electron with characteristic kinetic energies given by the reaction He* + Li → He + Li$^+$ + e$^-$. The ICD process of Li-doped He nanodroplets has been studied using synchrotron radiation and is discussed in detail elsewhere [46]. This process is generally an ultrafast and efficient decay mechanism in different weakly bound systems, such as rare gas clusters, hydrogen-bonded molecular complexes, and liquids [47]. An essential aspect of ICD is the production of low-energy electrons, which are genotoxic when created in living tissues [48].

The ions produced by ICD of resonantly excited Li-doped He nanodroplets are measured by TOF mass spectrometry. The inset of Fig. 4 shows TOF mass spectra recorded at a photon energy tuned to the peak of the 1s2p $^1$P resonance at 21.5 eV [49]. The mass spectrum recorded with an unfocused VUV beam is shown as a blue line, and the one recorded with a back-focused beam is represented as a grey-shaded area. When the unfocused beam interacts with Li-doped He nanodroplets, only Li ions of the two isotopes $^6$Li and $^7$Li, as well as $^7$Li$_2^+$ dimers produced by ICD are detected (Fig. 4 inset, blue line). The significant broadening of the peaks is explained by the large ionization volume of the unfocused VUV beam, which impairs the resolution of the mass spectrometer. No He ions are present, except for a tiny He$^+$ peak (4 amu), which is related to the presence of the 9$^{th}$ harmonic ($h\nu$=27.6 eV) in the unfocused beam. When the VUV beam is back-focused (and the 9$^{th}$ harmonic is suppressed by the multilayer mirror), additional sharp peaks appear in the mass spectrum. We attribute the appearance of He$_n^+$, $n$=1, 2, 3 ions to the collective autoionization process (CAI), which will be discussed in the following section. The $^7$Li$^+$ signal comprises a broad feature from the unfocused beam and a superimposed small sharp peak from the back-focused beam. As ICD from He* to Li is a process driven by the absorption of one photon per droplet, the integrated signal of the unfocused beam is much larger than that of the focused beam due to the larger interaction volume of the unfocused beam.

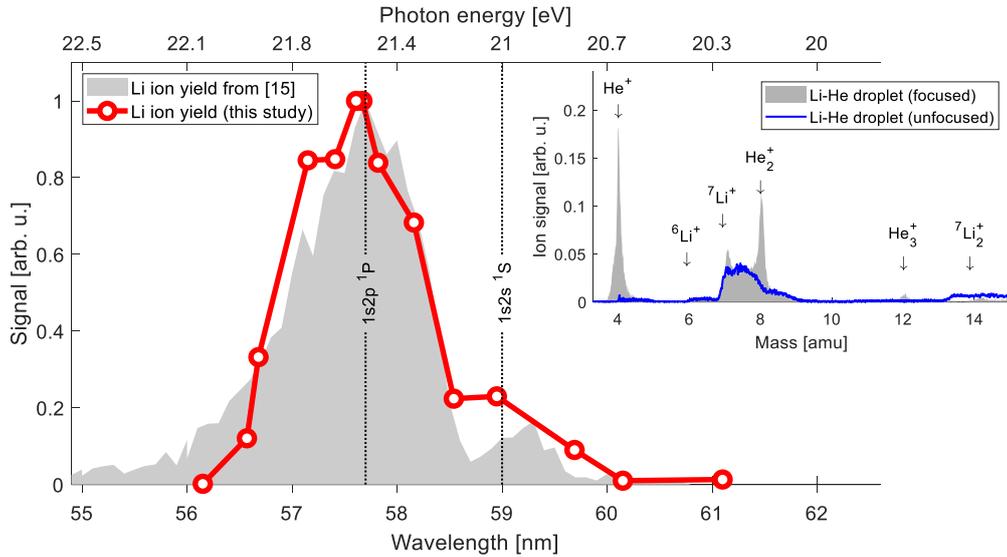

**Fig. 4 VUV absorption spectrum of He nanodroplets at the 1s2s/p band.** Li+ ion yield (in red) recorded from Li-doped He nanodroplets that are excited by a spectrally tuned 7$^{th}$ order harmonic radiation. The statistical variation of the signal at each data point is always lower than 12%; error bars are omitted for clarity. The grey shaded area is a reference Li+ ion yield spectrum recorded for Li-doped He nanodroplets by absorption of VUV photons from synchrotron radiation [15]. The vertical dashed lines indicate the energy of the nanodroplet He* 1s2p $^1$P and 1s2s $^1$S absorption lines. **Inset:** Ion yield spectra recorded for Li-doped He nanodroplets with focused (shaded) and unfocused (in blue) beam at h$\nu$ = 21.5 eV.

Here, we exploit the Li$^+$ ion yield generated by ICD as a sensitive probe for resonant excitation of He nanodroplets [15]. Fig. 4 shows the VUV photon energy-dependent integrated yield of Li ions produced by irradiation of Li-doped He nanodroplets by VUV pulses at photon energies tuned across the dominant optical absorption band of He nanodroplets in the range 20.0-22.0 eV in steps of 0.15 eV (red circles). The photon flux generated by our source is sufficiently strong even in an unfocused beam, so to avoid absorption of several photons by the same droplet, which would generate He2+ signal that overlaps with the 7Li+ signal, these measurements are performed with the unfocused beam by removing the back-focusing mirror. As described above, a small contribution from a distant 9th harmonic present in an unfocused beam is not relevant. The grey-shaded area shows the Li$^+$ ion yield measured by resonant excitation of Li-doped He nanodroplets with synchrotron radiation [15]. The main absorption band with a maximum around $h\nu$ = 21.5 eV is associated with the 1s2p $^1$P excited state of He. The side band around $h\nu$ = 21.0 eV is related to the lowest optically accessible 1s2s $^1$S state. The Li$^+$ ion signal measured with tunable VUV pulses using our HHG scheme closely follows the reference Li spectra from [15]. The results in Fig. 4 show that our HHG scheme provides a source of continuously tunable VUV radiation over a wide wavelength range without significant loss in efficiency. Note that the energy resolution of 0.15 eV is limited by the bandwidth of the 7$^{th}$ harmonic (see Fig. 1).

**Multiple excitations of He nanodroplets by intense monochromatic VUV pulses.** To assess the on-target VUV intensity achieved by tightly focusing the VUV beam, we have investigated the TOF ion mass and electron spectra of pure He nanodroplets. Fig. 5a shows a typical mass spectrum measured with the back-focused VUV beam at the absorption resonance of He nanodroplets, $hv = 21.5$ eV. He$_n^+$, $n=1, 2, 3$, mass peaks with similar strengths as for the case of Li-doped He droplets (shaded area in the inset of Fig. 4) indicate autoionization of the He nanodroplets following resonant excitation by the intense VUV pulses. Under our conditions, two or more He* excitations may occur in one nanodroplet, which can decay by the ICD-type reaction He* + He* → He + He$^+$ + e$_{ICD}$ [42, 50]. For sufficiently high VUV intensities inducing a large number of He* excitations per nanodroplet, the decay is dominated by a CAI mechanism involving many excited centers and various scattering processes. This process was previously observed for He nanodroplet irradiated by intense FEL pulses [51, 42].

More detailed information about the autoionization processes occurring in multiply excited He nanodroplets are obtained from electron VMIs and electron kinetic energy distributions extracted from the images using the Abel inversion method MEVELER [52]. Fig. 5b shows the background-subtracted electron velocity-map images recorded for pure He nanodroplets with 1200 He atoms per droplet. The upper image is taken with a focused VUV beam, and the lower one with an unfocused beam. For the unfocused beam, only one circular structure is visible, which is attributed to photoelectrons directly released upon absorption of photons from the 9$^{th}$ harmonic component. The non-homogeneous center-symmetric structure of the images is caused by the reduced sensitivity of the phosphor screen. Photoelectrons from the 9$^{th}$ harmonic are also present in the image recorded with the focused beam (upper image) with equal intensity. In addition, a larger ring and a central bright spot are visible in the upper image. To identify these features, we inspect the electron spectrum generated from the difference between the two images, shown as a red line in Fig. 5c. The peak at about 16.6 eV results from the ICD reaction mentioned above, which generates an ICD electron with an energy of 16.6 eV assuming that the two He* excitations are in their 1s2s $^1$S atomic states [50]. The sharp feature close to zero electron kinetic energy is indicative of electron emission of thermalized multiply excited He nanodroplets which evolve into nanoplasmas by CAI [42]. The grey shaded area shows an electron spectrum previously measured at the FEL FERMI with a beam intensity of about $10^{10}$ W/cm$^2$, three times longer pulses, and at a similar He droplet size (2500 He atoms per droplet) [42]. The appearance of a CAI signal in our electron spectrum with similar intensity as in the FEL experiment confirms the high VUV intensity on a target of about $3 \times 10^{10}$ W/cm$^2$. We note that the intensity level estimated from comparing these two data sets is more than one order of magnitude lower than the maximum intensity estimated from our experimental parameters (see Methods). We assume that this discrepancy is mainly due to the very short Rayleigh length of the focused VUV beam (10 μm) compared to the diameter of the nanodroplet beam (4 mm). Consequently, the effective intensity contributing to the measured signal is reduced by averaging over the focal volume. The CAI signal strongly depends on the size of the nanodroplets because it is a many-body process that relies on the absorption of many photons by a nanodroplet [42]. As seen from the inset of Fig. 5c, it rises approximately linearly by a factor of 2 in the given range of nanodroplet sizes.

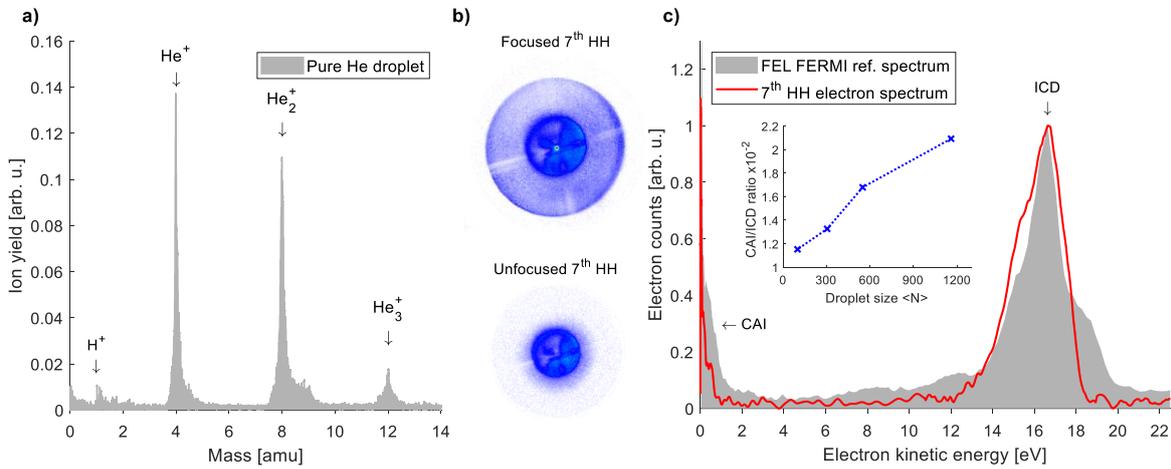

**Fig. 5 Ion mass and electron spectra of He nanodroplets irradiated by intense resonant VUV pulses. a.** Ion mass spectrum for pure He nanodroplets with back-focused VUV beam. **b.** Electron velocity-map images recorded for pure He nanodroplets of size $N \approx 1200$ with back-focused and unfocused VUV beam (7$^{th}$ harmonic) tuned to the photon energy $hv = 21.5$ eV. **c.** Electron spectrum inferred from the difference between VMI's for back-focused and unfocused beams (red line). For reference, the grey-shaded area shows an electron spectrum measured with VUV pulses generated by the free-electron laser FERMI, taken from [42]. **Inset:** The ratio of collective autoionization process versus interatomic Coulombic decay integrated signals measured for variable He nanodroplet sizes.

**Discussion**

We have demonstrated a technique to produce tunable monochromatic femtosecond VUV pulses with high intensity based on broad spectrum of our high-power 1 kHz OPCPA laser system. Our approach preserves a relatively high conversion efficiency throughout the tuning range. It allows us to achieve an intensity on target sufficient to drive non-linear collective interactions in doped He nanodroplets.

The tuning range for a particular harmonic is limited mainly by the available bandwidth of the laser system. It is defined by the system design and the broadband optics' properties, and especially the high dispersion chirped mirrors for pulse compression. Our present system operates in the 750-930 nm range, but this could be extended with further development of the laser source and optics. In our work, we have focused on tunability in the 20 eV range, where other tuning techniques fail, but we expect our approach to be easily extended into different spectral ranges.

The spectral resolution is mainly defined by choosing the appropriate thickness of the non-linear conversion crystal. Increasing the crystal thickness will narrow the bandwidth of the generated high harmonics. Therefore, the SH driving scheme allows for adjustment of the spectral resolution of the setup (e.g., as in Fig. 4) according to the application's needs.

Using the SHG drive results in a high conversion efficiency in the HHG process, thereby increasing the photon flux at the desired photon energy by almost two orders of magnitude according to wavelength scaling laws [53, 30]. This is due to the lower dispersion of the electron wave packet driven by a field with a shorter period. With proper geometry setup, the shorter wavelength of the HHG driver also leads to a reduced VUV beam focal spot size of a single back-reflecting ML mirror as the HHG source diameter is downsized. Future upgrades of the laser system [43] delivering a pulse energy of 100 mJ or beyond, will allow an increase in photon flux by up to one order of magnitude according to the scaling laws [54].

The capabilities of our continuously tunable VUV source are shown in a proof-of-principle experiment on ICD in doped He nanodroplets. However, the potential of the setup extends to a wide range of applications where spectral tunability is crucial for exciting specific electronic transitions and accessing resonances. Prospective applications of tunable VUV light sources include new developments in coherent diffractive imaging, where rich features in the diffracted images are expected to arise upon resonant sample excitation [13], monitoring correlated electron dynamics in heterogeneous nanosystems [14, 55] and new regimes of photochemical processes [1, 56]. Thus, the realization of a spectrally tunable HHG source with high conversion efficiency and low-loss VUV beam propagation, where the pulse properties stay primarily unaffected by optical elements (like a grating monochromator), is highly desirable for applications.

The high VUV intensity on the target was verified by observing characteristic features in the electron spectra of He nanodroplets which are indicative of multiple excitations leading to binary autoionization by ICD and even to CAI by many-body interactions. We benchmark our present results to previous FEL studies and confirm that the central results are reproduced when the yields of electrons generated by ICD and CAI were measured systematically for variable He droplet sizes. The linear increase of the CAI electron yield relative to that of ICD electrons reflects the higher absorption cross-section of larger droplets leading to an increased probability of absorption of many photons. CAI becomes the dominant decay mechanism when the fraction of excited atoms in a He droplet exceeds about 1% [44].

The presented source can be employed in a pump-probe setup to perform time-resolved studies of ultrafast phenomena in atoms, molecules, or extended systems [57, 58], which require precise spectral tunability and high photon flux. As such, it represents a tabletop alternative to seeded FEL facilities for various applications.

The instruments described in this publication are available to the international user community through an open-access application procedure.

## Methods

**Tuning of the OPCPA.** The laser system used for the experiment consists of 5 OPCPA amplification stages, each pumped by 3 ps pump pulses at 515 nm. For this particular experiment, the delay of the pump was adjusted relative to the seed providing more energy in the portion of the spectrum used at the moment. In addition, the compression in the spectral range of interest was optimized using a Dazzler, which allowed us to achieve the desired shape of the SHG spectrum.

**Second harmonic generation.** A 200 micron-thick BBO crystal type I cut at the angle of 29.2° was placed 220 cm from the focus of a spherical mirror with a focal length of 5 m (f-number of 150). The estimated laser intensity on the frequency-doubling crystal was 1 TW/cm$^2$. The conversion efficiency across the tuning range was estimated to be between 30%-40%, with a predicted SH pulse duration shorter than 30 fs. Optimal conversion efficiency at the desired central wavelength of SH was achieved by adjusting the OPCPA system as described above.

**VUV beam characterization.** Spectral characterization of the generated VUV beam used for the He resonances measurement is done at the beginning and the end of the measurements for each data point to ensure the beam properties remain the same. For this, we have used our wavelength-calibrated VUV flat-field spectrometer that can be inserted into the beam under vacuum to keep the same stable generation conditions when switching between characterization and application. As an online method to check the proper UV pump setting during the measurements, we have implemented an online UV-NIR spectrometer collecting the residual pump beam from a reflection of the metallic aluminum filters used for separating the VUV light. The reflection was collected by a collimator and transported by an optical fiber for characterization by a UV-NIR spectrometer HR4000 (OceanInsight). By this method, we ensure that the spectral properties of the UV beam remain the same during the experiments, indirectly verifying the stability of the VUV during the measurement process.

**Calibration of VUV spectrometer.** The VUV flat-field spectrometer is wavelength-calibrated through the measurement of high-harmonic spectra driven by IR laser. A set of specific metallic filters (Al and Mg) was used for calibration. The calibration has been performed by observing the spectral absorption edges and their higher diffraction orders. By this method, we obtained a set of defined photon energies used as fit parameters to obtain a precise VUV spectrometer calibration.

**Conversion efficiency and VUV intensity on the target.** For the calibration of the photon flux delivered on the target, we placed a VUV CCD camera in the direct beam in the position of the back focusing mirror (see Fig. 2). By this method, we obtained the VUV beam profile on the focusing mirror. From the calibration of the camera signal, we deduced the number of photons from all harmonics transmitted through the 200 nm thick Al filter. The number of photons in the 7$^{th}$ harmonic is then deduced from the measured VUV spectrum. Using the HHG source distance, the measured beam profile on the back-focusing mirror, and the geometrical shape of the mirror, we have performed the physical optics propagation simulation obtaining an ideal focal spot size of about 1μm diameter at 1/e$^2$. VUV pulse duration is estimated from the high-harmonic bandwidth to be always shorter than 30 fs. The number of photons is corrected to the reflectivity of a broadband multilayer back-focusing mirror with a high reflectivity coating up to 27 eV. From that, we obtained an estimated value of $2\times10^7$ photons/shot delivered on the target at a 1 kHz repetition rate. In combination with the estimated focal spot size, the maximal reachable peak intensity on the target was about $7\times10^{11}$ W/cm$^2$. Note that the Rayleigh range is only around 10 μm; thus, the VUV intensity averaged over the interaction region with He droplets is lower. Estimating the transmission of the used aluminum metallic filter, we can calculate the conversion efficiency in the source from IR to the 7$^{th}$ harmonic to be of the order of 10$^{-5}$.

**Generation of He nanodroplets.** A beam of pure He nanodroplets was produced by the expansion of high-pressure He gas (50 bar) through a pulsed (repetition rate = 50 Hz), cryogenically cooled Even-Lavie nozzle with an inner diameter of 80 μm. The generated He nanodroplet beam was then propagated through a first 1 mm skimmer, a vacuum chamber for doping purpose, and a second 3 mm skimmer before crossing the VUV beam in the VMI spectrometer. By varying the expansion conditions (backing pressure and nozzle temperature), the mean droplet size is varied in the range of <N> = 100-1500 He atoms per droplet. Titration measurements determined the He nanodroplet sizes following the procedure reported in [59]. To measure the Li$^+$ ion signal generated by ICD, the He nanodroplets were doped with Li atoms evaporated from a heated cell with a length of 10 mm, as in [15]. The doping level of Li is adjusted slightly lower than the maximum probability for doping He nanodroplets with single atoms by setting the temperature of the heated cell to 360 °C.

**Electron and ion detection.** Electrons or ions coming from the VUV-irradiated He nanodroplets were detected on a single-shot basis (at 50 Hz) using a VMI spectrometer (Photek, VID275). The maximum voltage applied to the electrodes of the VMI spectrometer allowed the detection of electrons with kinetic energy up to 28 eV. The signal related to He-droplets was measured with the opening time of the Even-Lavie valve temporarily synchronized to the arrival time of the VUV pulses. The background signal due to photoionization of the residual gas and He and Li atoms effusing into the MAC chamber was measured by delaying the opening period of the Even-Lavie valve with respect to the arrival time of the VUV pulses. For each experimental run, 3000 single-shot TOF traces and velocity map images are recorded and then averaged. All ion and electron signals recorded as a function of the wavelength of the 7$^{th}$ harmonic are normalized in each step to the photon flux measured by a CCD camera placed directly at the position of the back-focusing mirror.

**Data availability.**
All relevant data are available on request from the corresponding author.

**Acknowledgments**
This work was supported by the project "Advanced research using high-intensity laser-produced photons and particles" (ADONIS) (CZ.02.1.01/0.0/0.0/16 019/0000789) from the European Regional Development Fund and the Ministry of Education, Youth and Sports. We acknowledge ELI Beamlines, Dolni Brezany, Czech republic, for the provided beamtime and thank the facility staff for their assistance and the Institute of Physics of the Czech Academy of Sciences for their support.
This project has received funding from the European Union's Horizon 2020 research and innovation programme under grant agreement No 871161.

**Author contributions**
M.K., M.M., J.N. and J.A. conceived the project, L.J. designed tunable VUV source, L.J., O.H., O.F., M.A., J.N. contributed to the development and operation of the HHG beamline, R.A., A.G., A.Š., W.S. contributed to the development and operation of the L1 laser, A.H.R., Z.H., E.K., M.K. and J.A. contributed to the development and operation of the MAC station, L.J., L.B.L., A.H.R., Z.H., E.K., A.S., M.M. performed the experiment, L.J., L.B.L., A.H.R., A.S., M.M. analyzed the data, L.J. wrote the manuscript with contributions from L.B.L., R.A., M.K., M.M., J.N., J.A. All authors reviewed the manuscript.